\documentclass[iop,twocolumn,apj]{emulateapj}
\usepackage{graphicx}
\usepackage{amsmath}

\newcommand{\feh} {\mbox{\rm [Fe/H]}}

\newcommand{\logg}  {\mbox{{\rm log}$g$}}
\newcommand{\gmr} {\mbox{$(g-r)_{\rm 0}$}}
\newcommand{\Mg} {\mbox{$M_{\rm g}$}}

\shorttitle{The Metallicity Gradient of the Thick Disk}
\shortauthors{Chen et al.}

\begin{document}

\title{The Metallicity Gradient of the Thick Disk Based on Red Horizontal Branch Stars From SDSS DR8}

\author{Y.Q. Chen, G. Zhao, K. Carrell, \& J.K. Zhao}

\altaffiltext{}{Key Laboratory of Optical Astronomy, National Astronomical Observatories, Chinese Academy of Sciences, Beijing, 100012, China; cyq@bao.ac.cn.}

\begin{abstract}
Based on SDSS DR8, we have selected a sample of 1728 red horizontal branch stars
with $|Z|<3$ kpc
by using a color-metallicity relation and stellar parameters.
The sample stars clearly trace a typical
thick disk population with peaks at $|Z|=1.26$ kpc and $\feh=-0.54$.
The vertical metallicity gradient of the thick disk is estimated in
two ways. One is a fit to the Gaussian peaks of the metallicity histograms
of the thick disk by subtracting minor contributions from the thin disk and the inner halo
based on the Besan\c con Galaxy model. The resulting gradient is
$-0.12 \pm 0.01$ dex kpc$^{-1}$ for  $0.5 <|Z| <3$ kpc.
The other method is to linearly fit the data based on stars
with $1 <|Z| <3$ kpc being the main component of the thick disk.
Five subgroups are then selected in different directions in the $X-|Z|$ plane
to investigate the difference in the vertical metallicity gradient between
the Galactocenter and anti-Galactocenter directions. We found that a
vertical gradient of $-0.22 \pm 0.07$ dex kpc$^{-1}$ is detected for
five directions except for one involving the pollution of stars from
the bulge. The results indicate that the vertical gradient is dominant,
but a radial gradient has a minor contribution for the thick disk population
represented by RHB stars with $1 <|Z| <3$ kpc.
The present work strongly suggests the existence of a metallicity gradient
in the thick disk, which is thought to be negligible in most previous works
in the literature.

\end{abstract}

\section{Introduction}
\label{sec:intro}
The presence of the thick disk component in the Galaxy has been
established for several decades \citep{gil83}. The age, metallicity
and rotational velocity of thick disk stars have been investigated for
various types of stars in the solar neighborhood. However, the formation
mechanism of the thick disk is not well known. Actually, the metallicity
gradient for the thick disk, being a basic input for the model of the Galactic disk,
is not well studied in the literature.
To the contrary, the traditional radial metallicity gradient of the Galactic thin disk
is well investigated based on H$II$ regions, Cepheids, OB stars,
open clusters (young population) and planetary nebulae (old population).
It is found that the radial gradients vary with time in the
sense of a steeper slope in the early stages compared to the present value,
and there is a change in slope at large Galactocentric distances $R > 10$ kpc
\citep{Maciel09}.

 Generally, the thick
disk is typical for vertical distances from the Galactic plane
of 0.5 to 1.5 kpc based on observational data in the solar neighborhood. With a large
spectroscopic survey of stars far away from the Sun, it is
suggested that stars with $1 <|Z| <3$ kpc dominate the thick
disk \citep{Allende06}, and the edge of the thick disk could
reach up to $|Z| \sim 5.5$ kpc. The large range in the $|Z|$
distribution enables us to investigate the vertical metallicity
gradient of the thick disk, which is not possible for the thin disk
where most stars are located within $|Z| < 0.5$ kpc. Such a study
has recently begun and
the existence of a metallicity gradient in the thick disk
is a discrepant issue. For example, \citet{Allende06} proposed no metallicity
gradient in the thick disk for $ 1 <|Z| <3$ kpc with a upper limit
on the gradient of 0.03 dex kpc$^{-1}$, while
\citet{katz11} investigated the metallicity distribution function (hereafter MDF) of the thin-thick-disk-halo system
in two Galactic directions
for several intervals between 0 and 5 kpc above the Galactic plane and
detected a vertical metallicity gradient of $-0.068 \pm 0.009$ dex kpc$^{-1}$.
\citet{Siegel09} suggested that the thick disk presents no metallicity gradient.
But some models with a gradient are compatible with observations of
the thick disk dominated by a population with $1 <|Z| <4$ kpc as pointed
out by \citet{katz11}.
Meanwhile, in an indirect way,
the mean metallicity for the thick disk varies with different heights
from $-0.48 \pm 0.05$
at $0.4 <|Z| <0.8$ kpc \citep{Soubiran03}, to $-0.685 \pm 0.004$
and $-0.780$ at $1 <|Z| <3$ kpc \citep{Siegel09}, which may indicate
the existence of a metallicity gradient. Note that these works
are mainly based on dwarf stars and they have limitations in some
aspects. For example,
\citet{Soubiran03} have high resolution but low signal-to-noise
data for low heights of $0.4 <|Z| <0.8$ kpc where both the thin and thick disks
contribute significantly to the MDF. \citet{Allende06} and \citet{Ive08} were
based on an SDSS DR6 sample
where metallicity on the metal rich side could be underestimated by 0.2-0.3 dex
according to \citet{Bond10}. \citet{Siegel09} are based on photometrically derived
metallicities, which are not as good as spectroscopically derived values.
Recently, \citet{katz11} provide high quality data for the metallicity
estimation, but the number of sample stars is limited to several hundreds.

The aim of this work is to probe the metallicity
gradient of the thick disk
population using red horizontal branch (hereafter RHB) stars selected from SDSS DR8.
This study has some advantages over previous studies.
First, the number of stars in
the sample is on the order of thousands, and
the metallicity is based on the updated version in SDSS DR8.
Second, RHB stars are an interesting type of star, which are
different from dwarf stars, with stellar distances
easily obtained from their intrinsic absolute magnitude.
Their high brightness make them traceable to distances of
up to 10 kpc.
Finally,
we investigate the metallicity distributions for
thick disk stars in five different directions in the $X-|Z|$ plane.
In particular, the comparison of the thick disk properties between
the Galactocenter and anti-Galactocenter directions has great potential
for providing more stringent data for model
comparisons. Thanks to the large survey area of the SDSS project, it is
the first time that such properties in different directions can be studied
and they will provide important clues for our understanding
of the thick disk formation.

\section{The Sample Stars}

The selection procedure of RHB stars is similar to \citet{chen10}, but uses
SDSS DR8 \citep{aba11} rather than SDSS DR7 \citep{aba09}.
Specifically, the first selection criterion is the color-metallicity relation
of $\gmr = 0.343 (\pm0.039) \feh+0.829$ presented in \citet{chen10} with
a deviation in $\gmr$ to the above relation $\delta < 0.20\ mag$.
Then stars with $g_0<20\,mag$, $\feh > -2.0$ and signal-to-noise ratios
of their spectra larger than 10 are used to obtain good quality data.
Thirdly, RHB stars should have stellar parameters within the range
of 4500--5900\,K in temperature and 1.8--3.5 dex in $\logg$.
In the selection, photometric
data and stellar parameters (including metallicities) in photometric
and spectroscopic STAR tables of DR8 are adopted.
Stellar distances of stars are calculated from $g$ magnitudes and
the absolute
magnitude-metallicity relation
of $\Mg=0.492\feh+1.39$ derived by \citet{chen09} with metallicities
taken from the SDSS DR8 spectroscopic STAR table. The
stellar distance and position on the sky is then converted to
the cartesian system of Galactic coordinates (X,Y,Z).

 As shown in \citet{aba11}, the photometric
data and stellar parameters are updated in DR8 along with an enlarged survey
area in the disk. Both improvements provide a better chance for us to probe
the thick disk properties.
With the aim to investigate the properties of the thick disk,
the sample stars are selected to have $|Z|< 6$ kpc since
the edge of the thick disk is at
about 5.5 kpc above the Galactic plane as suggested by \citet{Majewski94}.
With this limit, we extract a sample of 2,729 RHB candidate stars
from SDSS DR8.
The left panel of Fig.~1 shows the distributions of distance to the Sun, vertical distance to
the Galactic plane and Galactic latitude of the sample.
In particular, the $|Z|$ distribution of RHB stars has a peak around $|Z|=1.2$ kpc
and most of them are located within $|b|=6-16^\circ$ with a weak tail extending to $|b|=40^\circ$.
There is a hint of a minimum in the $|Z|$ distribution
at $|Z|=3$, which is taken as the selection limit of the thick disk in many works.
After that, $|Z|$ flattens out from $|Z|=3$ kpc to $|Z|=6$ kpc
where we cut our sample.
As we know that many
previous works select thick disk stars with $|Z|<3$ kpc or $|Z|<4$ kpc,
in the present work, we limit our sample to $|Z|<3$ kpc and
1728 RHB stars are selected for studying the metallicity gradient.
Most of them are limited to within 10 kpc and $|b|<20^\circ$ as shown
in the dashed lines in the left panel of Fig.~1.

\begin{figure}[bt]
\begin{center}
\includegraphics[scale=1.00]{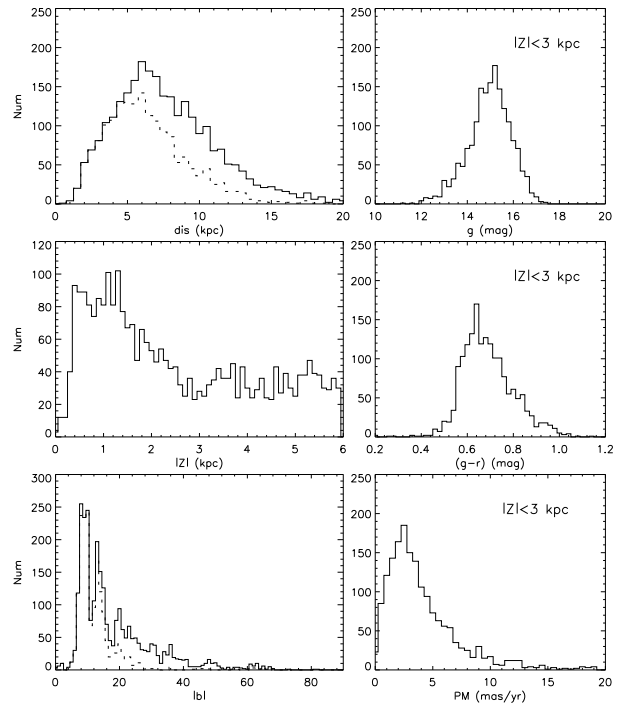}
\caption{Left: The distributions of distance, $|Z|$ magnitude
and galactic latitude for RHB stars with $|Z|<6$ (the thick line)
and $|Z|<3$ (the dashed line). Right: The $g_0$ magnitude, $\gmr$ color
and proper motion distributions for stars with $|Z|<3$. \label{fig:f1}}
\end{center}
\end{figure}

The right panel of Fig.~1 shows the $g_0$ magnitude, $\gmr$ color
and proper motion (PM)
distributions of our sample stars. In comparison with the selection 
criteria for stars in the SDSS/SEGUE survey \citet{Yanny09},
we found the main contribution of our sample stars comes from K (and
partly red K) giants, which covers the metallicity range of $\feh \sim -2.0$ 
to the solar value (with $l^e > 0.07$), has $g$ magnitudes brighter 
than 18.5 $mag$, $PM < 11\ mas/yr$ and a color range of $\gmr = 0.5$ to $0.8\ mag$. 
These criteria will not produce a significant
distance or metallicity bias in our sample selection, and thus
we assume that these target
selections in SDSS/SEGUE will not affect our result.

\section{Analysis and Results}
\subsection{The metallicity distribution of the thick disk}
The metallicity distribution of our sample stars is shown in the upper panel of
Fig.~2. It shows that the contribution from the thin disk is significant
with 550 stars having $\feh>-0.25$ and a peak at $\feh=-0.12$.
The peak at $|Z|=0.8$ kpc in Fig.~1 partly corresponds to this population.
The halo has a negligible contribution with only 77 stars having
$\feh<-1.0$. There are 1101 stars with $-1.0 < \feh <-0.25$, corresponding
to the thick disk, which have peaks at $|Z|=1.26$ kpc and $\feh=-0.54$.

In order to exclude the contribution of the thin disk and
the halo, we obtain simulation data from the Besan\c con
Galaxy model \citep{Robin03} using the selection criteria:
(i) $12 < g < 20\ mag$; (ii) $b = -30^\circ$ to $30^\circ$ with a step size of $10^\circ$; 
(iii) $l = 50^\circ$ to $250^\circ$ with a step size of
$25^\circ$; (iv) absolute magnitude $M_{V} = 0.0 -1.2$ and
spectral type from F to K; (v) population being either thin disk or halo;
and (vi) the color-metallicity relation and stellar
parameter ranges for selecting RHB stars in \citet{chen10}. All these
criteria fit our observed RHB star sample.
Then, the metallicity distributions of the simulated data for
the thin disk and halo are normalized to the star numbers
in the observed RHB sample for each population respectively.
Coincidently, the simulated thin disk from the model has the same
metallicity peak at $\feh = -0.115$ and the simulated halo has
negligible contribution to the metallicity distribution as
shown by the dashed lines in the upper panel of Fig.~2. The MDF of the thick disk
after subtracting the simulated thin disk and halo contributions
is shown in the lower panel of Fig.~2. A Gaussian fit to the
metallicity histogram is performed and has a peak at
$\feh =-0.54$, which is quite close to the peak of the histogram.

\begin{figure}[bt]
\begin{center}
\includegraphics[scale=1.00]{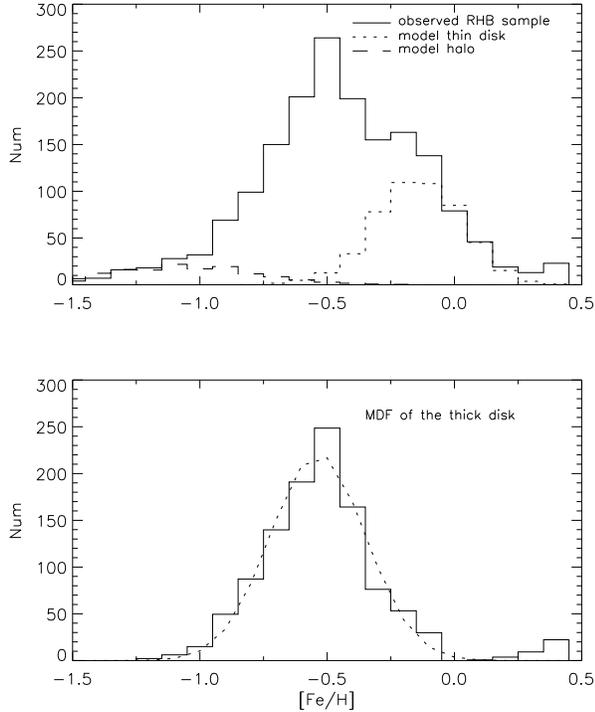}
\caption{The $\feh$ distribution of the sample stars before and
after subtracting
the contributions from the model thin disk and halo populations.}
\end{center}
\end{figure}

\subsection{The metallicity gradient of the thick disk}
In our sample, there are significant numbers of stars at different
$|Z|$ ranges of $0.5-1.0$ kpc, $1.0-1.5$ kpc, $1.5-2.0$ kpc, $2.0-2.5$ kpc, and $2.5-3.0$ kpc
in the middle panel of Fig.~1. Thus, it provides a way to investigate
the metallicity gradient for $0 < |Z| < 3$ kpc.
We perform the same procedure to each of the five different $|Z|$ bins as
that in Sect.3.1
and obtain the Gaussian peak of the metallicity histogram for each $|Z|$ bin.
Fig.~3 shows the metallicity distribution for the $1.0 < |Z| < 1.5$ kpc bin before and after subtracting
contributions from the thin disk and halo components based on model data.
The resulting peak metallicities from the Gaussian fits to the metallicity
histograms for five different $|Z|$ bins are shown
in Table 1, as well as the height range, the median $|Z|$ and the star counts.
The bin size for the plotted histograms are 0.05 dex and we have checked
that a bin size of either 0.02 or 0.10 will not significantly change the
Gaussian peak in the fit, but the distribution will deviate from a Gaussian
function for one or two bins.
The errors in the metallicity peak in Table 1 are estimated by the
Gaussian dispersion of the metallicity peaks from
1000 simulated samples randomly selected from the observed samples
for each height bin using a bootstrap method.
Fig.~4 shows the peak metallicity of the thick disk MDF as a function
of median $|Z|$ and the metallicity gradient is estimated
from the weighted least squares fit to the data. The metallicity
gradient is $-0.12\pm0.01$ dex kpc$^{-1}$  with an intercept of $-0.34$ dex.
This gradient is slightly steeper than that of \citet{katz11} who found
$-0.068 \pm 0.009$ dex kpc$^{-1}$ with an intercept of $-0.46$ dex.

In order to estimate the errors from the Besan\c con
Galaxy model, we vary the normalization factors in both
the halo and the thin disk components by reducing and enhancing
the adopted value by 10\%, which is a worst case scenario, and the modeled
metallicity distributions deviate clearly from the
observed distributions. We therefore obtain an additional four sets of Gaussian
metallicity peaks of the thick disk for each bin. In total, we have
five sets of metallicity distributions for five $|Z|$ bins, which produces
3125 metallicity gradients in the lower panel of Fig.~4. All
these slopes lie within $-0.11$ to $-0.15$, and thus the errors
from uncertainties in the  Besan\c con model do not have a significant
effect on the metallicity gradients.

\begin{table}
\caption{Thick disk metallicity peak as a function of height}
\begin{tabular}{rrrr}
\hline
Height & median Z & TD peak  & Num\\
 kpc   & kpc      &          & \\
\hline
$[0.5,1.0]$ & 0.74 & $-0.432 \pm 0.017$ & 416 \\
$[1.0,1.5]$ & 1.26 & $-0.480 \pm 0.013$ & 443 \\
$[1.5,2.0]$ & 1.74 & $-0.559 \pm 0.017$ & 308 \\
$[2.0,2.5]$ & 2.23 & $-0.620 \pm 0.019$ & 242 \\
$[2.5,3.0]$ & 2.71 & $-0.651 \pm 0.033$ & 159 \\
\hline
\end{tabular}
\end{table}

\begin{figure}[bt]
\begin{center}
\includegraphics[scale=1.00]{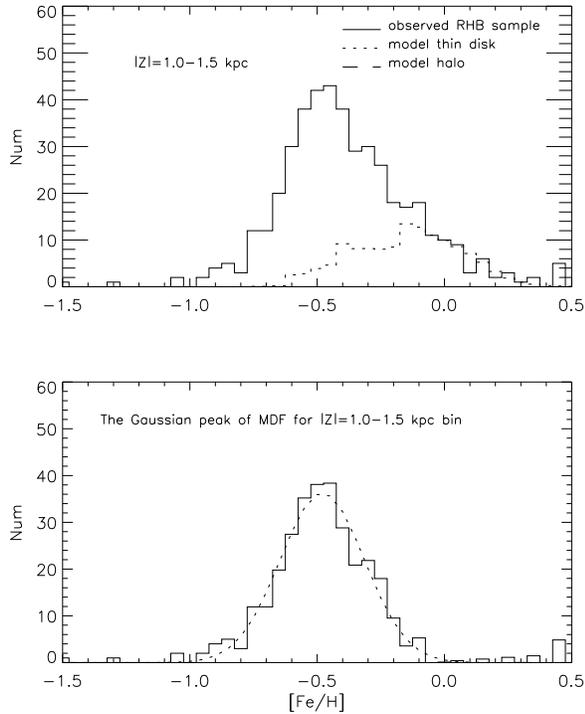}
\caption{The same as Fig.2 but for the $1.0 <|Z| < 1.5$ bin.}
\end{center}
\end{figure}

\begin{figure}[bt]
\begin{center}
\includegraphics[scale=1.00]{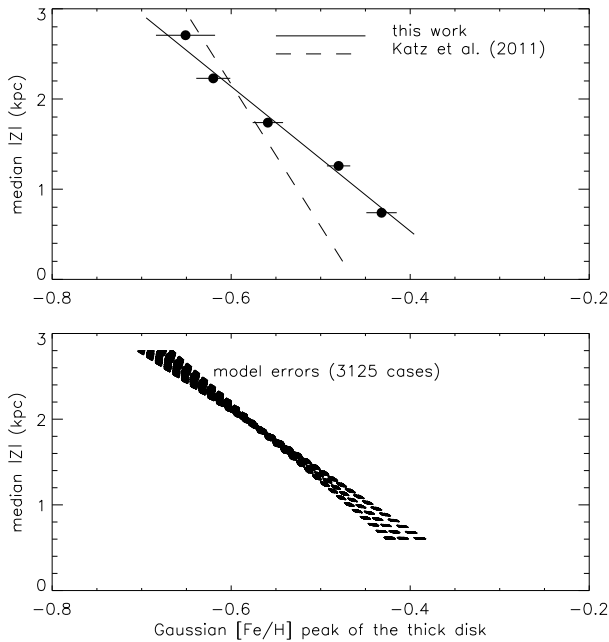}
\caption{The metallicity gradient for the thick disk for $|Z|=0-3$ kpc in
this work (solid line) as compared with that from \citet{katz11} (dashed line).}
\end{center}
\end{figure}

\subsection{The $\feh$ versus $|Z|$ diagram in five different directions in the $X-|Z|$ plane}
\citet{katz11} compared the metallicity distributions in two different
directions in order to investigate the origin of the gradient from either
vertical or radial variations. They found that the vertical origin is strongly
favored and the Kolmogorov-Smirnov test (hereafter KS test)  shows similar metallicity distributions for two
fields directed at M3 and M5 with different $X-Z$ directions.
Since both of the two directions in \citet{katz11} are located in the anti-Galactocenter
direction, it is interesting to make this comparison for different directions,
including some Galactocenter directions.
Meanwhile, the previous method to estimate the metallicity gradient is based on
the assumption that the thick disk does not include stars with high metallicity
($\feh \sim -0.12$), which is arbitrarily assigned to be the thin disk population, and
does not include stars with low metallicity ($\feh < -1.0$), which is arbitrarily
assigned to be the halo population. Actually, the thick disk can extend to a metallicity
of $\feh=0.0$ and a metal weak component of the thick disk is widely accepted
to exist in the Galaxy. Therefore, we perform an alternative way to trace
the metallicity gradient using the $\feh$ versus $|Z|$ diagram, which is
used in many works.
In particular, it is important to know how the metallicity distributions
vary between the Galactocenter and anti-Galactocenter directions since
the scale length of the thick disk typically ranges from 2.2 to 3.6 kpc \citep{Morrison90,Robin96,
Carollo10}.
In our sample, it is possible to select stars in five different directions
in order to investigate the properties of the thick disk
on both sides of the solar circle. Fig.~5 shows the selection of five groups (G1 to G5)
of stars in the $X-|Z|$ plane with the G1 and G2 fields toward
the Galactic center, G4 and G5 toward the anti-Galactocenter direction,
and G3 is limited to the solar region of $X=8.3$ to 9.3 kpc.
Here X is along the line
between the Sun and the Galactic center and $|Z|$ is the absolute value
of the vertical distance to
the Galactic plane.

\begin{figure}[bt]
\begin{center}
\includegraphics[scale=1.00]{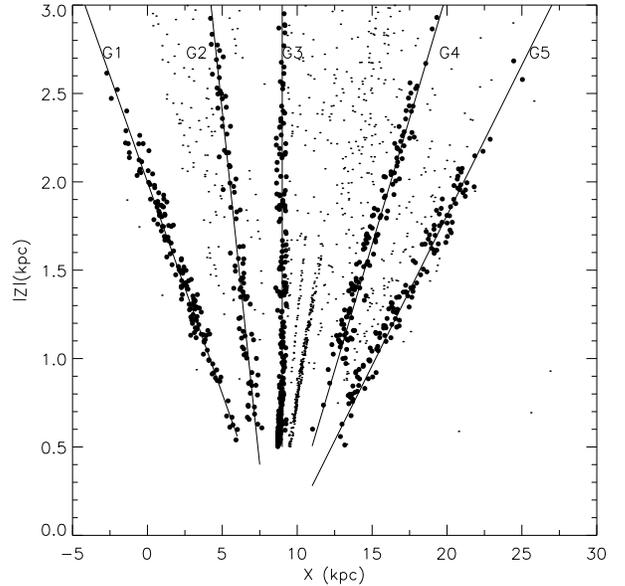}
\caption{Five star groups, G1 to G5, selected in the $X-|Z|$ plane.\label{fig:f5}}
\end{center}
\end{figure}

Fig.~6 shows the $\feh - |Z|$ diagrams for all stars with
$0.5 <|Z| < 3.0$ kpc in five directions. Stars with $|Z| <0.5$ kpc
are excluded because they are missing in most directions and large
contributions from the thin disk persist.
The metallicity gradients are investigated by using mean metallicity at
different height bins for five groups of stars.
It is clear that all five groups show the existence of a metallicity
gradient. A careful inspection
of Fig.~6 shows the first two points with $|Z| < 1.0$ kpc are quite
different inside the solar circle (G1 and G2)
compared to outside the solar circle (G4 and G5).
In view of the fact that the thick disk is predominant for $1< |Z| <3$
kpc while a star located between 0.5 and 1.0 kpc has equal chances to be in either the
thin or thick disk populations according to the Besan\c con model of the Galaxy
\citep{Robin03}, it is more reasonable to estimate the metallicity gradient
 among the different
groups only for stars with $1< |Z| <3$ kpc.
Assuming that the variation of $\feh$ with $X$ can be neglected (see discussions in Sect. 3.4), the vertical
metallicity gradients can be estimated with a linear fit to the data using the
formula $\feh=a|Z|+b$,
which usually provides a higher
slope than that from the peak of the distribution at each bin.
For comparison, we derive such metallicity gradients for five groups,
which is given in Table 2.
The metallicity gradients for the four groups G2-G5 are similar
with a slope of $-0.22$, but have decreasing intercepts from 0.02 (G2) to $-0.20$ (G5).
The regression to G1 gives a slope of $-0.367$ and an intercept of 0.07.
A bootstrap method is used to estimate the errors of the
slopes by adopting 1000 simulated samples randomly selected
from the observed five groups.  The results are shown in Fig.~7.

\begin{table}
\caption{The metallicity gradient in the formula
of $\feh = a\, |Z|+b$ based on data for the five groups.}
\begin{tabular}{rrrrr}
\hline
Groups & a  ($\sigma_a$) & b &  scatter & Num\\
\hline
G1 & $  -0.367\pm0.074$ & $  0.078$ &   0.162 &140\\
G2 & $  -0.228\pm0.069$ & $  0.019$ &   0.148& 123\\
G3 & $  -0.217\pm0.058$ & $ -0.057$ &   0.199& 259\\
G4 & $  -0.227\pm0.040$ & $ -0.102$ &   0.110& 125\\
G5 & $  -0.225\pm0.060$ & $ -0.191$ &   0.196 & 113\\
\hline
\end{tabular}
\end{table}

\begin{figure}[bt]
\begin{center}
\includegraphics[scale=1.00]{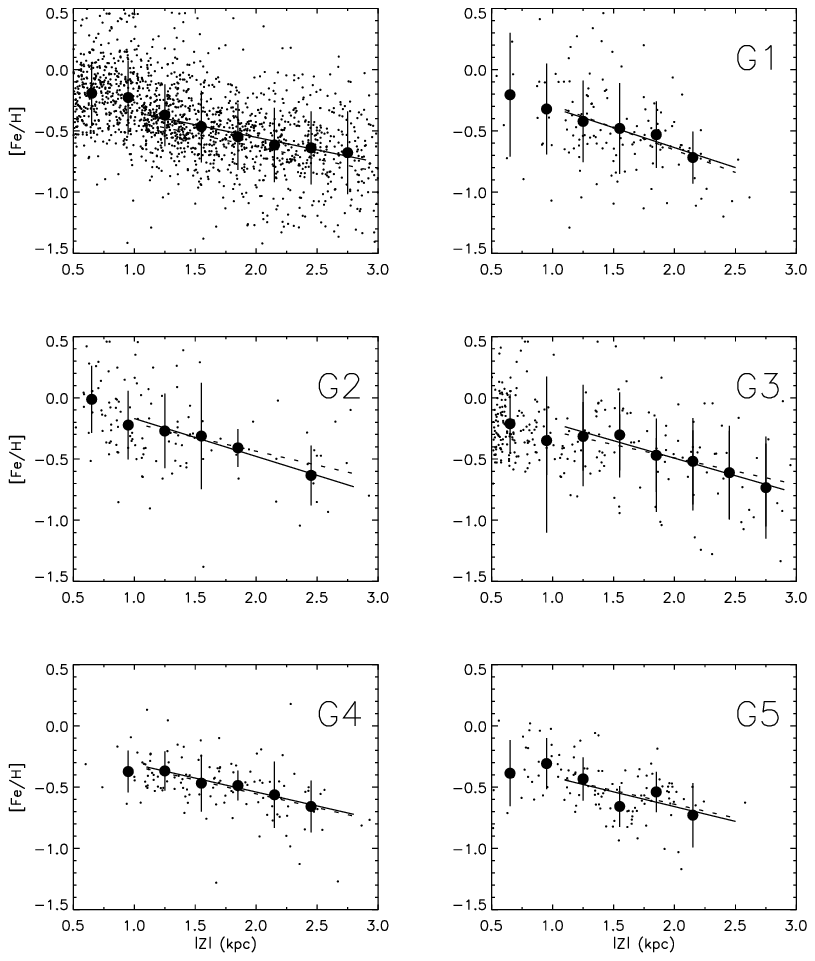}
\caption{The five subsamples in the $\feh - |Z|$ plane.\label{fig:f6}}
\end{center}
\end{figure}

\begin{figure}[bt]
\begin{center}
\includegraphics[scale=1.00]{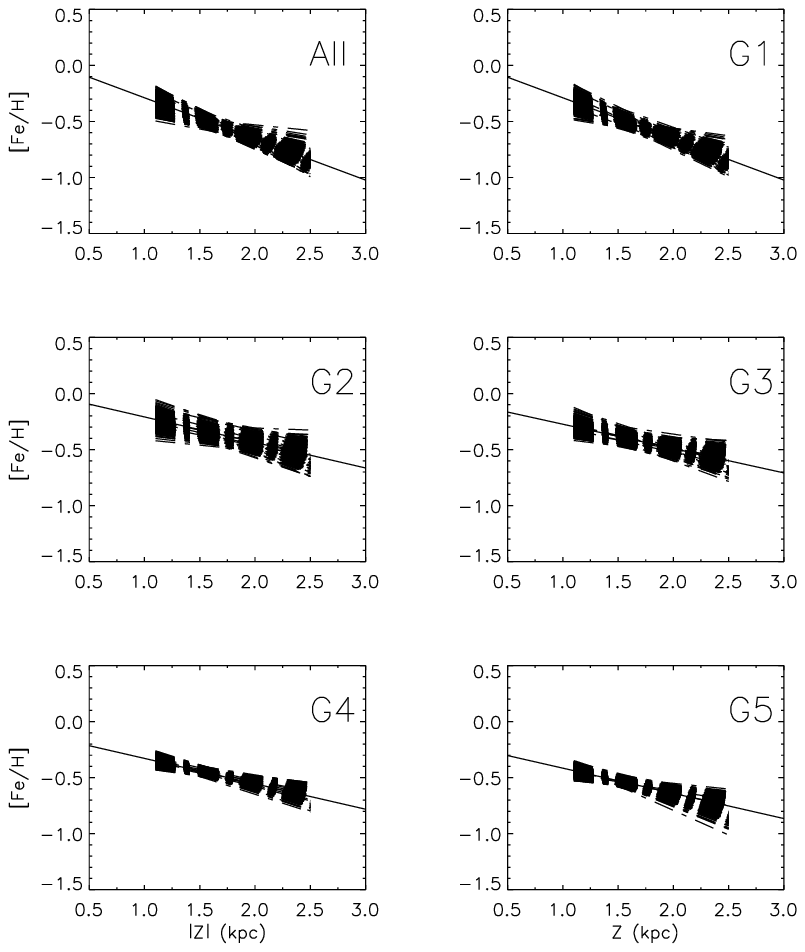}
\caption{The five groups in the $\feh - |Z|$ plane.\label{fig:f7}}
\end{center}
\end{figure}

\subsection{Vertical versus radial gradients}
When we fit linear relations to $\feh$ as a function of $|Z|$, the
variation of $\feh$ with X is assumed to be negligible. This may be the
case for G2 to G5. Note that G2 to G5
have different X variations: G3 has nearly a fixed X value and G2 has little
X variation, while G4 and G5 have significant X variation. Similar slopes in the
$\feh$ versus $|Z|$ fits for G2 to G5 indicate that the variation of $\feh$
with X can be disregarded in the $\feh$ versus $|Z|$ fits. This does not
mean that there is no radial (hereafter X is referred to as radial) metallicity
gradient. Instead,
the decreasing intercept ($b$ in Table 2) from G2 to G5 indicates a negative
radial gradient and a linear fit to the $b$ versus median X for G2 to G5
gives a slope of $-0.013$ dex kpc$^{-1}$.
Such a small slope has no significant effect on the $\feh$ versus $|Z|$ fits 
for our G2 to G5 where X variations are less than 5 kpc.

The small effect of X variation on the vertical metallicity gradient
for G2 to G5 can be further confirmed by similar metallicity distributions
among these groups with different median X values for similar $|Z|$ ranges
of $1<|Z|<3$ kpc. For this purpose,
we investigate the cumulative distribution function
(hereafter CDF) for $\feh$ via the KS test among the five groups.
For the two groups in the anti-Galactocenter direction G4 and G5,
similar $\feh$ CDFs with
a $p-$value of 0.14 are shown in Fig.~8.
This agrees with \citet{katz11}
where the two directions toward M3 and M5 in the anti-Galactocenter region
show similar $\feh$ distributions. Since G4 is located inside of G5, similar
$\feh$ CDFs indicate the variation of metallicity with X is not detectable.
Similarly, G2 and G3 have similar $|Z|$ distributions with little or no 
X variation within the group, and a KS test yields a $p-$value of 0.33 for the 
$\feh$ CDFs based on stars
with $1< |Z| <3$ kpc. Again, G2 is located inside of G3 and similar CDFs
indicate X differences between G2 and G3 do not have a significant effect on
their metallicity distributions.
Similar $\feh$ CDFs are found for G3 and G4, but
this is not the case for the comparison between G1 and G5.
Fig.~9 shows the similar $|Z|$ but different $\feh$ CDFs between
G1 and G5, being the two extreme cases in opposite $X$ directions.
For $-0.7 < \feh < -0.4$, G1 and G5 have exactly the same CDFs
while at both ends the differences in the $\feh$ CDFs are significant.
Two possible reasons may explain the special case of G1.
It has been suggested that the effect of $\feh$ variation with X
cannot be neglected for a scale length of the thick disk between 2 and 4 kpc. 
Alternatively,
it is expected that G1 has a high contribution of stars at both
the metal poor end and the metal rich end from
the bulge population
with a wide metallicity range of $-2.0 < \feh < 0.5$,
which will steepen the gradient.
In addition, it seems that G1 also has a special kinematic distribution
in Fig.~10, which shows the radial velocity distributions for five groups with 
G1 having the largest radial velocity range and a peak significantly
different from the other groups. Therefore, G1 is not included in
this study.

\begin{figure}[bt]
\begin{center}
\includegraphics[scale=1.00]{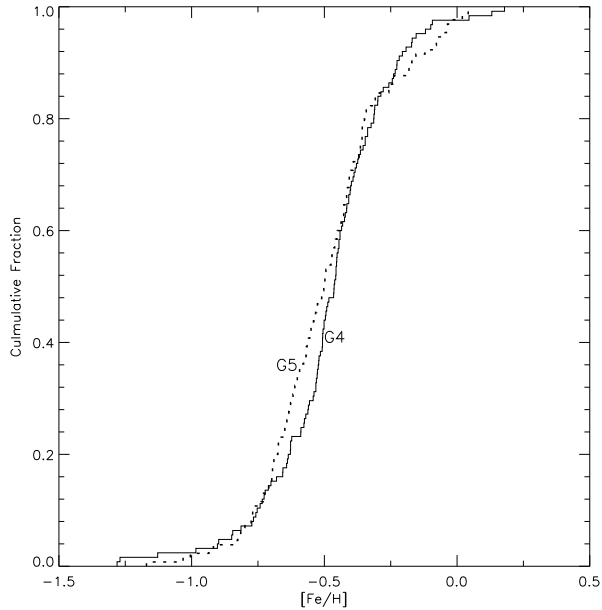}
\caption{The cumulative distributions of G4 and G5 by KS test.\label{fig:f8}}
\end{center}
\end{figure}

\begin{figure}[bt]
\begin{center}
\includegraphics[scale=1.00]{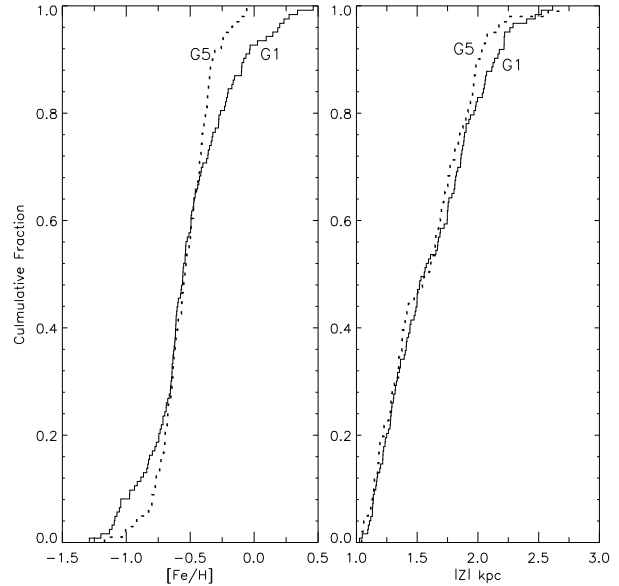}
\caption{Comparison of the $\feh$ distribution between G1 and G5 by KS test
for $1< |Z| <3$ kpc.\label{fig:f9}}
\end{center}
\end{figure}

\begin{figure}[bt]
\begin{center}
\includegraphics[scale=1.00]{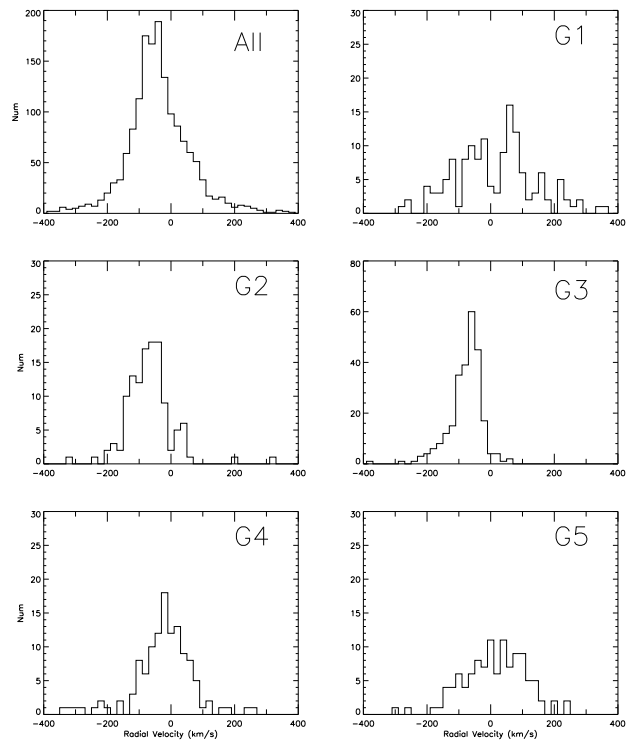}
\caption{Comparison of the radial velocity distribution for five groups
at $1< |Z| <3$ kpc.\label{fig:f10}}
\end{center}
\end{figure}

\section{Summary}
We have estimated the metallicity gradient for the thick disk population
in two ways using RHB stars selected from SDSS DR8 data. The first
method is based on the Gaussian peak of the metallicity distribution in five
$|Z|$ bins by subtracting the contribution from the thin disk and halo via
the Besan\c con Galaxy model. The slope is $-0.12\pm0.01$ dex kpc$^{-1}$
 with an intercept
of $-0.34$ dex. The second method is linearly fitting the data in
the $\feh - |Z|$ diagrams directly
for stars from $1< |Z| <3$ kpc, where the contribution from the
thin disk and halo is not significant. Five groups in different directions
can be separated to test for consistency, and we found that they give similar
gradients of around $-0.225\pm0.07$ dex kpc$^{-1}$ in four groups, 
with the only exception for the direction with $-4 < X <4$. Both methods indicate
the existence of a metallicity gradient in the thick disk.
In our opinion, the first method gives a lower gradient because
the model predictions suppress the existence of both metal rich and metal poor
components in the thick disk.  We therefore favor the second method with a steeper
gradient because it comes from the data directly and
thus does not depend on the model assumptions.

\acknowledgments
This work has been supported by the National Natural Science Foundation of China
under grants No. 11073026, 10673015, 10821061, 11078019, the National Basic Research Program of
China (973 program) No. 2007CB815103/815403.

Funding for SDSS-III has been provided by the Alfred P. Sloan Foundation, the Participating Institutions, the National Science Foundation, and the U.S. Department of Energy Office of Science. The SDSS-III web site is http://www.sdss3.org/.
SDSS-III is managed by the Astrophysical Research Consortium for the Participating Institutions of the SDSS-III Collaboration including the University of Arizona, the Brazilian Participation Group, Brookhaven National Laboratory, University of Cambridge, University of Florida, the French Participation Group, the German Participation Group, the Instituto de Astrofisica de Canarias, the Michigan State/Notre Dame/JINA Participation Group, Johns Hopkins University, Lawrence Berkeley National Laboratory, Max Planck Institute for Astrophysics, New Mexico State University, New York University, Ohio State University, Pennsylvania State University, University of Portsmouth, Princeton University, the Spanish Participation Group, University of Tokyo, University of Utah, Vanderbilt University, University of Virginia, University of Washington, and Yale University.

\end{document}